# RECOVERING GAPS IN THE GAMMA-RAY LOGGING METHOD

**Ms. St. Nikita Churikov**[1]

**Assoc. Prof. Natalia Grafeeva**[2]

[1] Saint Petersburg State University, **Russia**
[2] Saint Petersburg State University, **Russia**

## ABSTRACT

The gamma-ray logging method is one of the mandatory well logging methods for geophysical exploration of wells. However, during the conduct of such a study, the sensor, for one reason or another, may stop recording observations in the well. If a small number of values are missing, you can restore these values using standard methods to fill in gaps like in time series. If data miss a large number of values, observations usually are made again, which leads to additional financial costs. This work proposes an alternative solution, in the form of filling missed observations in data with the help of machine learning methods. The main idea of this method is to construct a simple two-layer neural network that is trained on data from the well, and then synthesise the missing values based on the trained neural network. This work evaluates the effectiveness of the proposed method, and gives reasons for the appropriateness of using different methods of filling gaps, depending on the number of missed values.

## INTRODUCTION

Geophysical exploration of wells is an important part of the search for minerals in the earth. The process of exploring wells is long, time-consuming and requires serious financial investments. For this reason, when there is a failure in recording information about the well, researchers tend to try to come up with a way to avoid repeated measurements. A typical option is linear interpolation to fill in small spaces (up to 5 points), and larger gaps require more sophisticated methods or repletion of measurements, which leads to additional costs.

Also, the geophysicist, during the interpretation of the curve, could accidentally create small gaps. Visually, such omissions will be invisible, but numerically - these will be missing values.

After all, with automatic data processing and the use of time series, it is necessary to invent ways of processing passes, since algorithms can not work with them. The easiest way is to drop all values that are omissions, however this is a rather crude way, and we lose useful information using it. Therefore, there are various ways of filling the missing values in the data [1]. But for the most part, they are designed to use historical data and, as a rule, reduce to averaging based on the neighbourhood of the missing values [2]. Such methods are quite effective in case of small omissions (as mentioned above - up to 5 values), but absolutely not applicable in the case of a large number of missing data. Meanwhile, if other measurements were simultaneously performed, then it is possible to restore the missing values based on those measurements that were carried out in the well at the same time.

This paper is devoted to the evaluating of possibility of using machine learning methods (namely, a neural network) to restore gamma ray logging values based on historical data of other measurements.

## RELATED WORK

The problem of data recovery in wells is not new [3, 4]. Often, the articles consider the use of various classical machine learning algorithms to fill the gaps in the porosity, density, and permeability curves used to investigate the presence of oil in the well.

There is also a number of works devoted to identification of lithology. One of which is a result of machine learning competition [7]. This paper covers ways to preprocess well logging features and some ways to generate new data. But this dataset doesn't have missing values which is a vital problem in real world application.

During our initial research we've found papers on restoration of well logging methods stated above: porosity, density and permeability. That's why we've decided to focus our efforts on research of restoration of most commonly used method: gamma-ray well logging method as there are almost no articles devoted to its restoration.

## DATASET

For the purposes of the study, open data was used (http://doa.alaska.gov/ogc/DigLog/diglogindex.html). There are 53 wells in the dataset. In this paper, measurements were taken from 5 wells with a depth of more than 12k feet. In the source dataset, they have the following names: TRADING BAY UNIT D-05, D-06, D-09, D-12, D-13. In which the following logging methods were applied: GR, RHOB, SP, ILD, DT. In measurements corresponding to the readings of the sensors used in the gamma-ray logging method, a number of values have been missed, the number of which does not exceed half of the half the length of the well.

The missing values in the data are quite typical. They exist even in the data that lie in the public domain. In all used data-sets there are ~ 800 real gaps and the maximum number of consecutive missed values is 300, and on average 6 values are missing.

In general, there is a fairly large number of logging methods [5] and for specific tasks related to geophysics it is necessary to use certain methods. In the dataset, there are measurements corresponding to the following methods: GR, RHOB, SP, ILD, DT.

As the target variable we will consider GR, and as the observed values all other logging methods.

In Fig. 1 that most logs are similar to the normal distribution, but ILD has a heavy right tail, and SP has left and right values relative to zero and is asymmetric.

RHOB also has a similarity to a heavy left tail, but there are a lot of meanings, because this part can not be called a tail. rather, this behaviour of the curve as a whole.

Also, our target variable - GR is similar to the normal distribution, and so we can assume that the regression methods will restore it well.

## METHODOLOGY

The data presented is of a very different scale. So, for example, the SP method has values from -20 to -80, and DT ranges from 50 to 150. Also, the data is noisy, since it is obtained from sensors. It is very common in machine learning that data is to be prepared due to the the factors stated above. And in different knowledge area, different ways of preprocessing are required.

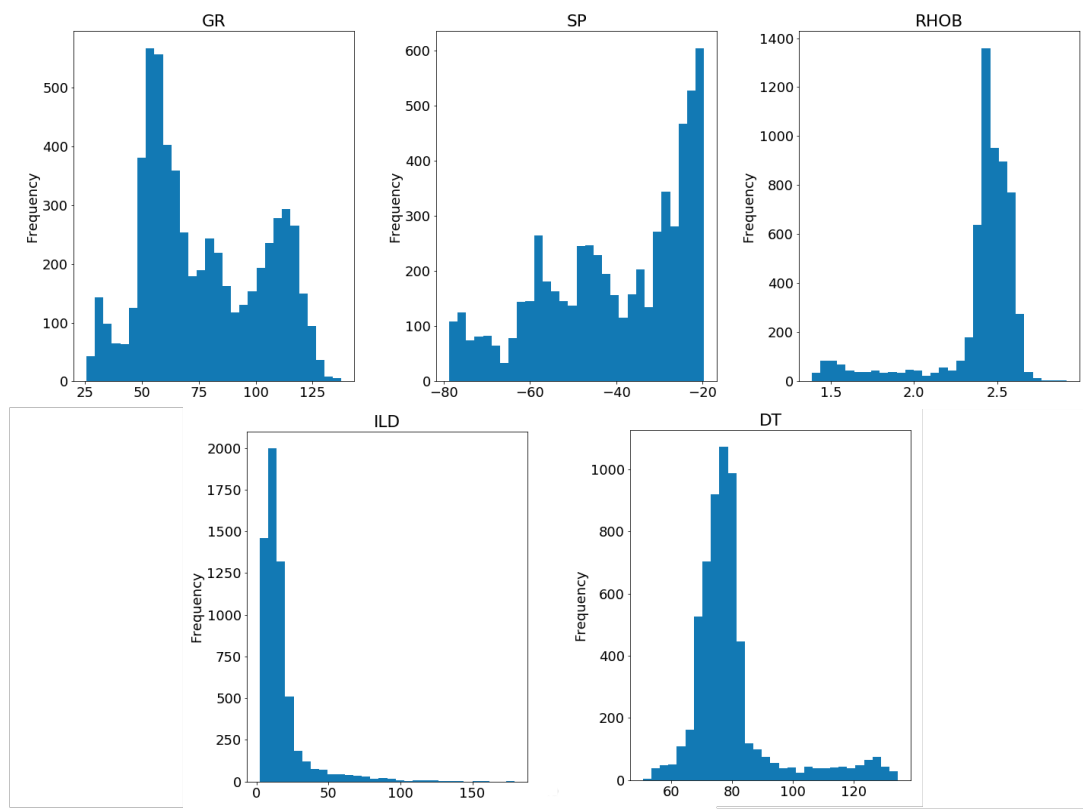

Fig. 1. Distribution of well logging measurements.

### *Preprocessing*

The following methods were used for preprocessing:

1. Standard Scaler to all data: since we transfer the data to a neural network, it is necessary that the data be on the same scale. The best standardizer was the Standard Scaler [6];

2. Apply *log(x+1)* to the ILD logging method: this transformation is a typical practice for

heavy tails;

3. To preprocess SP we used following pipeline: Direct Fourier transform -> Remove slow trends (We took the frequency 5) -> The inverse Fourier transform.

Idea of this transformation comes from observation that the deeper we go SP method begin to shift down: only the offset occurs. In order to get rid of this trend, you can apply the Fourier transformation transformation, trim low frequencies and then apply reverse Fourier transformation. As a result, the values will shift to the overall average trend as seen in the Fig 2. Green is the curve after we applied forward and backward transform, and blue is not preprocessed data.

### Generating new features

Since the data represent the discrete observations of the sensor at a certain depth, we came up with the idea that information should be added about what happened before and after the observation. So we decided to generate new features. There are various ways of generating new features [7].

In our study, we used several ways to generate new features using Quantile Smooth and Window generator:

*Quantile Smooth* - the idea of the method is that we take quantiles of some order not from the entire data set, but using the sliding windows from $k$ values before $i$-th observation and $k$ after. As a result, we get some new feature, which is a quantile of order n from a window of size $2k+1$ of the certain well logging method. Such a transformation can be perceived as smoothing out data. And logging curves are quite noisy, because local noise can be a problem.

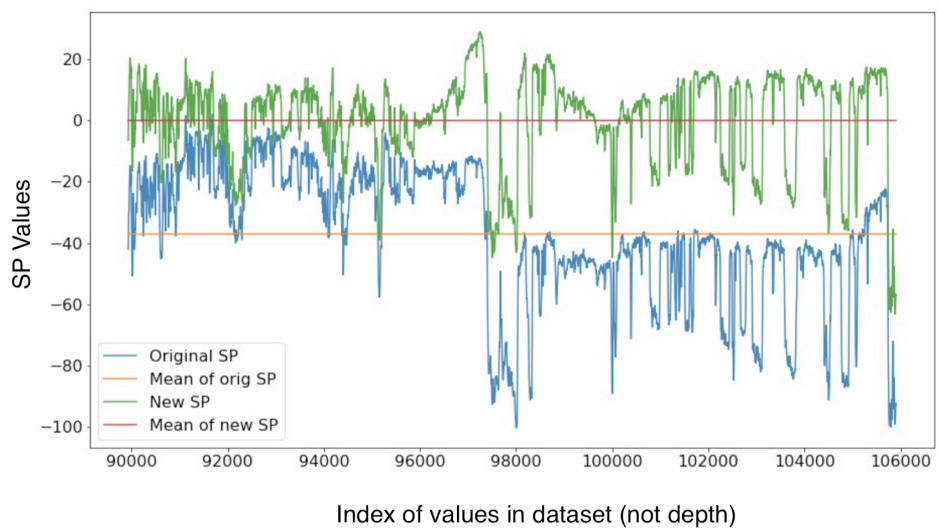

Fig 2. Before and after Fourier transform.

*Window generator* - Also the point in the log is a kind of convolution. So to unfold (deconvolve) it, we decided to take $n$ values of the curve up to some point $X_i$ and after it and create *2\*n* new features out of them. As a result, one observation will contain information about the data before and after observation. Such a transformation can be perceived as a sort of recurrent layer in a neural network.

### *Models*

The following models were used in the study:

• Linear interpolation

• Interpolation of the third order [8]

• Two-layer neural network [9, 10]

• Two-layer neural network with a "shift"

### *Linear interpolation*

This method is quite intuitive for small passes (up to 5 missed points). Its huge advantage in its simplicity: it does not require additional data, it only needs an initial and an end point to work in order to conduct a straight line.

### *Interpolation of third order*

Development of the idea with interpolation. In theory, a third-order polynomial is constructed and a somewhat more complicated curve is obtained. But, as will be shown later, this method does not work in this problem.

### *Two layer neural network*

We used a neural network of the following architecture: 100 neurons and batch normalization in the first layer, 50 neurons with batch normalization in the second layer.

We decided to use such a simple architecture for the purpose of fast learning on the data. At the moment, the network is being trained on the Quadro 4000 GPU for 1 minute. In total, with preprocessing and prediction of passes, it turns out about 2 minutes for learning and data recovery.

### *Two layer neural network with shift*

The specificity and advantage of our task is that we can add information from the training sample to the maximum. Often there are cases when the model understands the behaviour of a curve, but does not fall into the start and end points of the true curve. And at that moment one might want to kind of "move" predicted curve up or down into the right position. Linear interpolation has information about where to start and where to end it. Why shouldn't a neural network be given this information?

As a result, a combination of linear interpolation and a neural network led to the following idea:

• Train the model on the data present

• Use model to predict batch of the training data.

• Take the difference between the predicted learning and the true values and calculate the mean deviation

• Add it to the predicted values.

**EVALUATION**

To assess the quality, we used mean absolute error divided by the difference of 99% and 1% percentiles. This normalization allows you to place an error in the interval from 0 to 1 (the less, the better).

In Fig. 3 Translucent areas around the curves - the scatter of their values. The curves themselves are the mean values of the metrics at certain amount of missed values. One could see that for a sufficiently small gap size, linear interpolation works very well. However, with the increase in size, there is a need for more intelligent means. And NN with shift show good results of test data with small variance.

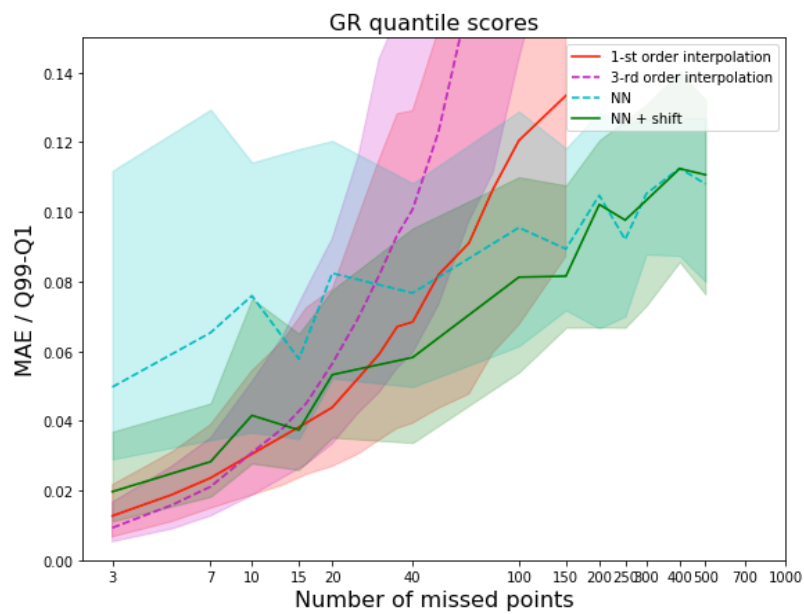

Fig. 3. Evaluation of predictions.

**CONCLUSION**

In the presented work, the application of neural networks to the recovery of gaps in the gamma-ray logging method was investigated. In the process of research, methods were found for preprocessing data, which allow to take into account information about neighboring observations. It has also been shown that for a small number of omissions it is sufficient to use simple methods such as linear interpolation, but for a large number of gaps it has been shown that using a neural network significantly reduces the error. The interpolation error grows exponentially, whereas the error of the neural network is linear. Also, a method for predicting omissions was found, which advantage of the fact that information can be taken from a training set.

The method obtained shows good results on a small number of omissions, and excellent results at large, which makes it a universal method for predicting omissions.

In a real application, one or another method could be used, depending on the number of omissions observed in the data.

**REFERENCES**

[1] Zhang Z. Missing data imputation: focusing on single imputation. *Annals of Translational Medicine*. 2016;4(1):9. doi:10.3978/j.issn.2305-5839.2015.12.38.

[2] Caruso, C. and Quarta, F., 1998, 'Interpolation methods comparison', Comput. Math. Appl. 35, 109–126.

[3] Lopes, Rui & Jorge, Alípio. (2017). Mind the Gap: A Well Log Data Analysis.

[4] Lopes, Rui & Jorge, Alípio. (2017). Assessment of predictive learning methods for the completion of gaps in well log data. Journal of Petroleum Science and Engineering. 10.1016/j.petrol.2017.11.019.

[5] Zorski, T., Jarzyna, J., Derkowski, A., Środoń, J. Well logging in the world of shale gas plays - Review of the logging methods [Geofizyka otworowa w dobie poszukiwań gazu w łupkach - Przegląd metod pomiarowych] (2013) Przeglad Geologiczny, 61 (7), pp. 424-434.

[6] Scikit-learn: Machine Learning in Python, Pedregosa et al., JMLR 12, pp. 2825-2830, 2011.

[7] Paolo Bestagini, Vincenzo Lipari, and Stefano Tubaro (2017) A machine learning approach to facies classification using well logs. SEG Technical Program Expanded Abstracts 2017: pp. 2137-2142.

[8] Sudhir Sharma, Er & Walia, Robin. (2013). Review Paper on Interpolation of Digital Images using Modal Function. International Journal of Research in Information Technology. 1. 85-90.

[9] Ioffe, S., Szegedy, C.Batch normalization: Accelerating deep network training by reducing internal covariate shift (2015) 32nd International Conference on Machine Learning, ICML 2015, 1, pp. 448-456.

[10] Jürgen Schmidhuber, Deep learning in neural networks: An overview, Neural Networks, Volume 61, 2015, Pages 85-117, ISSN 0893-6080